\documentclass[12pt]{article}

\usepackage{graphicx}

\begin{document}

\centerline{\bf Reexamination of scaling in the five-dimensional
Ising model}

\bigskip

Muneer A. Sumour, Physics Department, Al-Aqsa University, P.O.4051,
Gaza, Gaza Strip, Palestinian Authority, msumoor@alaqsa.edu.ps

\medskip

D. Stauffer,Institute for Theoretical Physics, Cologne University, 
D-50923 K\"oln, Euroland, stauffer@thp.uni-koeln.de

\medskip

M.M.Shabat, Physics Department, Islamic University, P.O.108, Gaza, 
Gaza Strip, Palestinian Authority, shabat@mail.iugaza.edu

\medskip

Ali H. El-Astal, Physics Department, Al-Aqsa University, P.O.4051, 
Gaza, Gaza Strip, Palestinian Authority, a-elastal@alaqsa.edu.ps

\bigskip

Abstract:

In three dimensions, or more generally, below the upper critical
dimension, scaling laws for critical phenomena seem well understood,
for both infinite and for finite systems. Above the upper critical
dimension of four, finite-size scaling is more difficult.

Chen and Dohm predicted deviation in the universality of the Binder cumulants for three dimensions and more for the Ising model. This deviation occurs if the critical point $T = T_c$ is approached along lines of constant $A = L^2(T-T_c)/T_c$, then different exponents a function of system size $L$ are found depending on whether this constant $A$ is taken as positive, zero, or negative. This effect was confirmed by Monte Carlo simulations with Glauber and Creutz kinetics.
 Because of the importance of this effect and the unclear situation in the analogous percolation problem, we here reexamine the five-dimensional Glauber kinetics.

\bigskip
\begin{figure}[hbt]
\begin{center}
\includegraphics[angle=-90,scale=0.5]{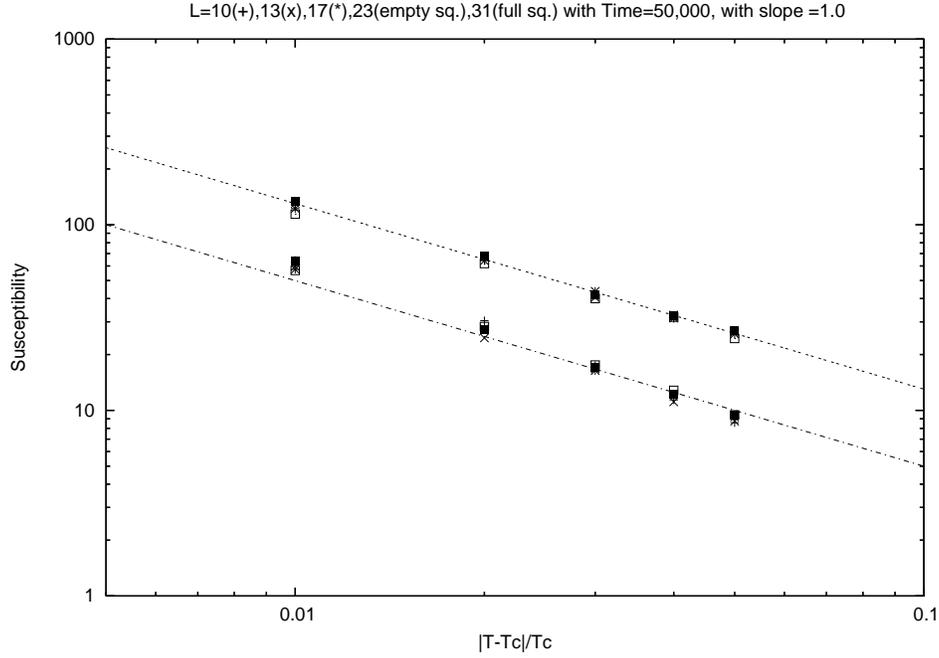}
\end{center}
\caption{Susceptibility versus temperature with different $L$ (=10, 13, 17, 23, 31), for 10 nearest neighbors as log-log plot, the upper data correspond to $T > T_c$ with amplitude 1.3 , and the lower to $T < T_c$ with amplitude 0.5 , and straight lines had the theoretical slope (-1).}
\end{figure}
\begin{figure}[hbt]
\begin{center}
\includegraphics[angle=-90,scale=0.5]{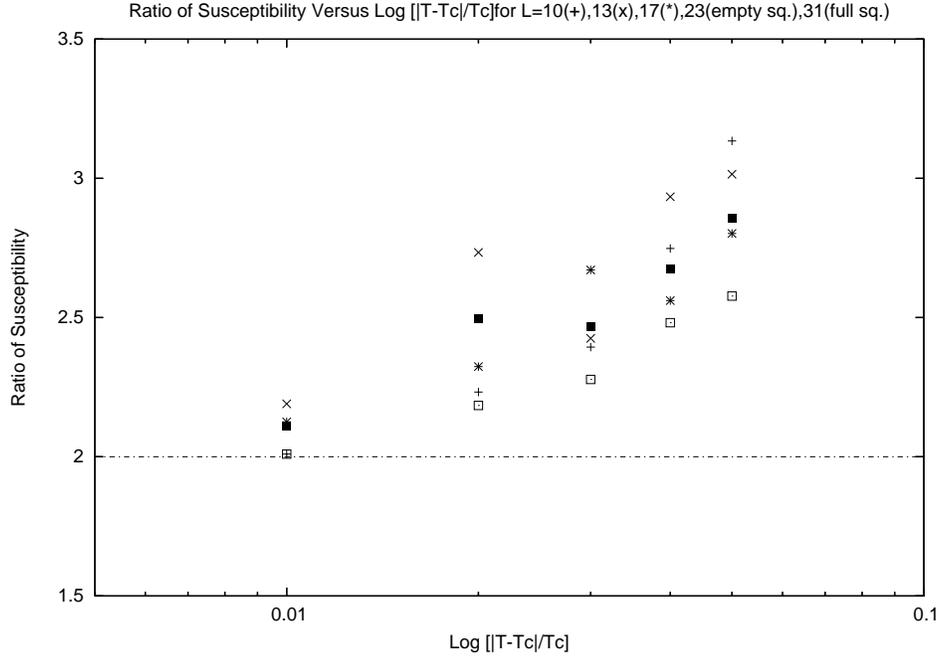}
\end{center}
\caption{Ratio of susceptibility above to below $T_c$, plotted semi-lo\-garithmically versus $|T_c-T|/T_c$, for  $L = $ 10,13,17,23,31, for 10 neighbors up to time = 50000.}
\end{figure}

\begin{figure}[hbt]
\begin{center}
\includegraphics[angle=-90, scale=0.5] {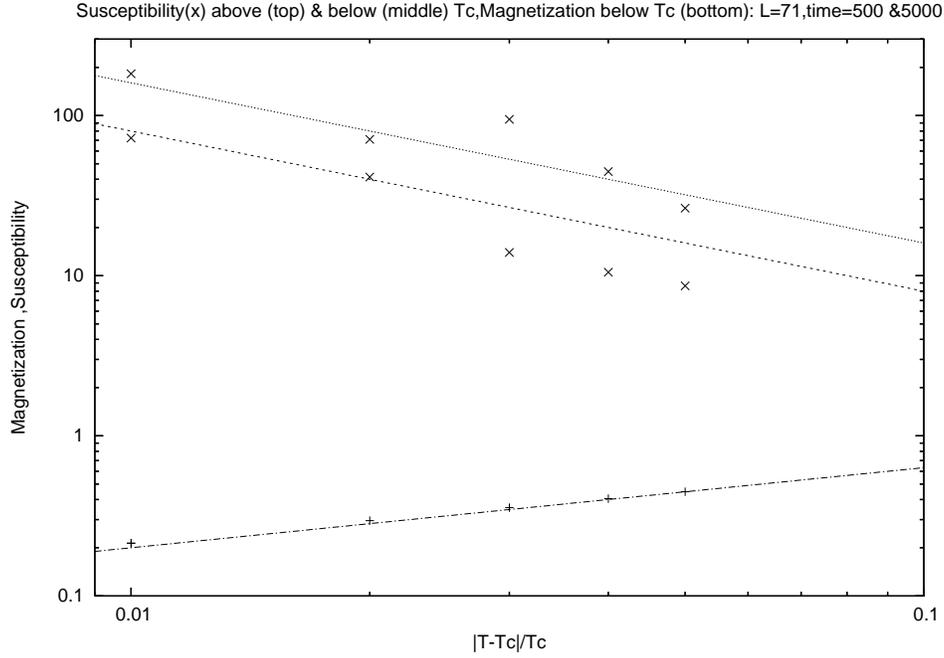}
\end{center}
\caption{$|M|$ and susceptibility versus $|T_c-T|/T_c$ with fixed size L
= 71 of lattice in log-log plot with lines indicating the
theoretical slopes -1 and + ½.}
\end{figure}

\begin{figure}[hbt]
\begin{center}
\includegraphics[angle=-90,scale=0.5]{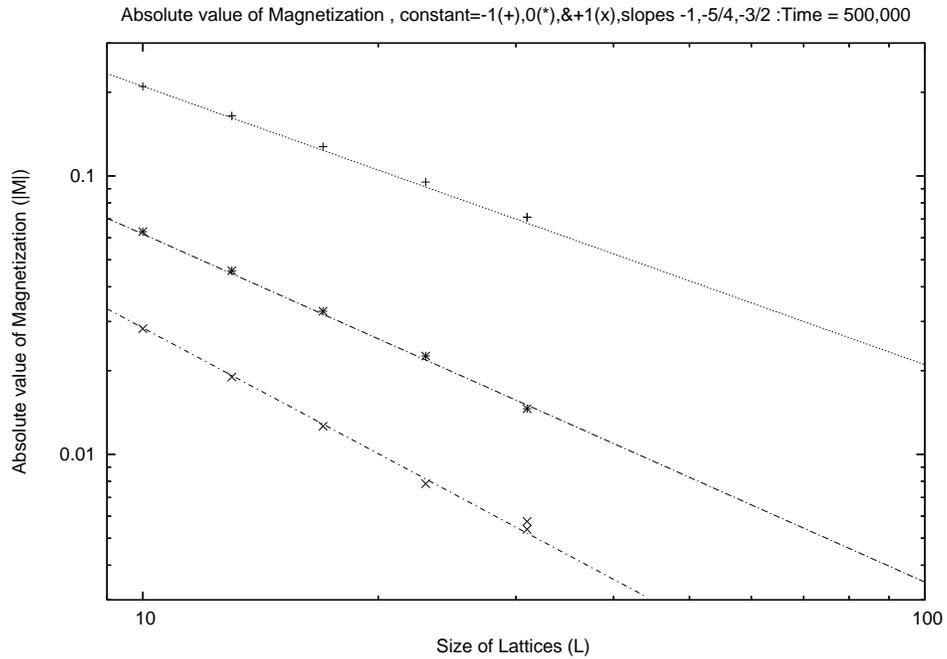}
\end{center}
\caption{$<|M|>$ versus size $L$ of lattice (10, 13, 17, 23, 31), in
log-log plot along constant $A = L^2(T-T_c)/T_c$. The upper data
correspond to $T<T_c\; (A=-1)$ with slope --1, the middle data to $T=T_c
\; (A = 0)$ with slope --5/4, and the lower to $T >T_c \; (A=+1)$ with slope
--3/2.}
\end{figure}

\begin{figure}[hbt]
\begin{center}
\includegraphics[angle=-90,scale=0.5]{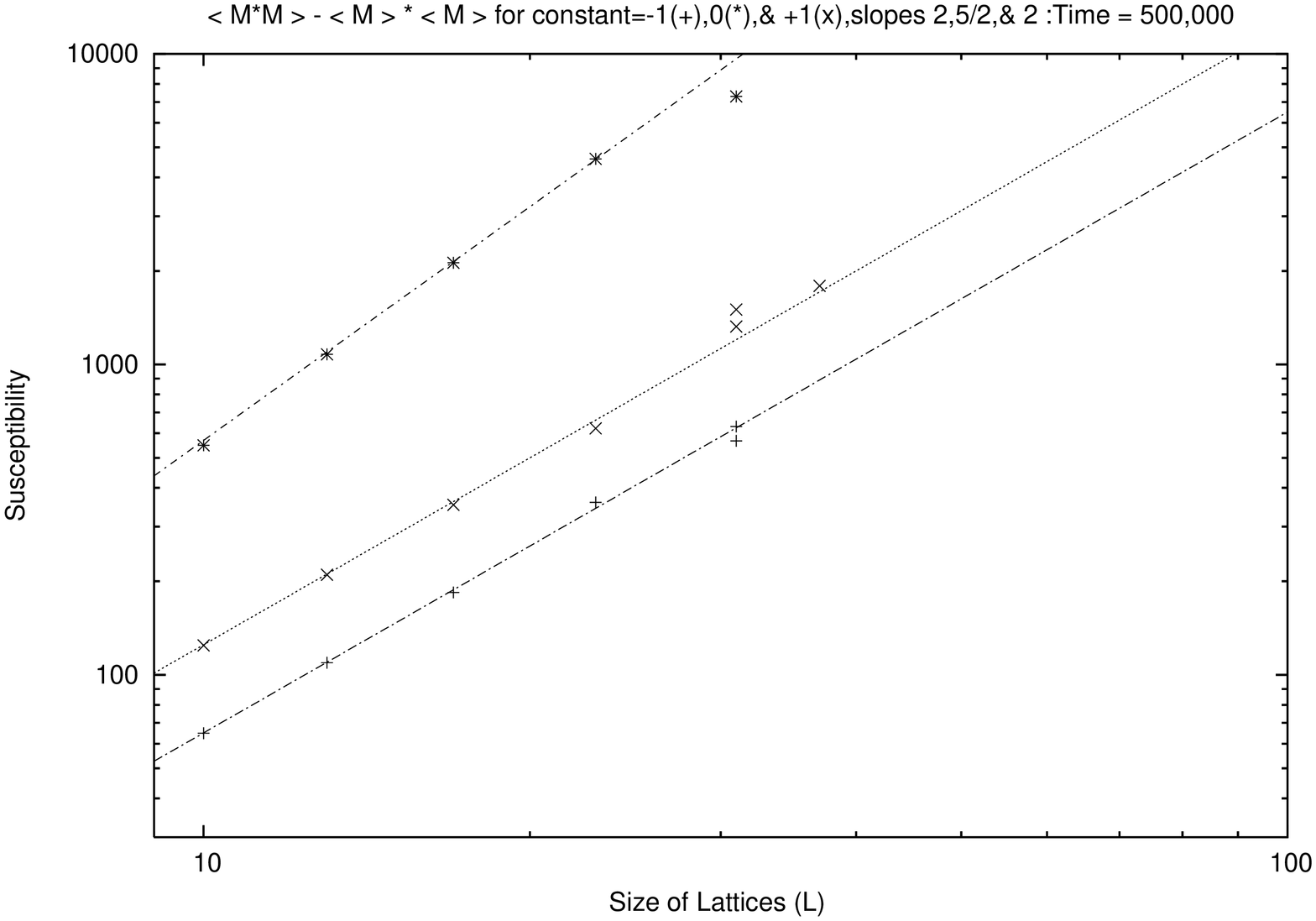}
\end{center}
\caption{ Susceptibility $<M^2> -<M>^2$ versus size $L$ of lattice
(10, 13, 17, 23, 31, (37)), as log-log plot along constant $A = L^2
(T-T_c)/T_c$, the upper data correspond to $T=T_c \; (A=0)$, the middle
data to $T > T_c \; (A=1.0)$, and the lower data to $T < T_c \; (A=-1)$. The
middle data fit better the indicated slope 2 than the expected slope
3.}
\end{figure}

1.
 {\bf Introduction}

In recent years, there is the question of universality of the
five-dimensional Ising model. This question focuses on the value of
susceptibility varying with temperatures near the critical
temperature for different sizes of lattices, and here we investigate
the susceptibility of the five-dimensional Ising model. In 2004 Chen
and Dohm predicted theoretically [1], and then this prediction was
partially confirmed [2-3], that the widely believed universality
principle is violated in the Ising model on the simple cubic
lattice with more than only six nearest neighbors. Other research
groups [4-7] studied the 2D and 3D Ising model for different
parameters and also for directed interactions problems occur in the
Ising model. Schulte and Drope [3] by Monte Carlo simulations with
Glauber [8] and Creutz [9] kinetics, found such violation, but not in
the predicted direction. Selke and Shchur [2] tested the square
lattice, for this importance effect and the unclear situation in the
analogous percolation [10], here we reexamine this universality for
the susceptibility ratio and magnetization near the critical point.
For this purpose we study first the standard 5D Ising model with ten
nearest neighbors.
Our study is based on Monte Carlo simulations for systems with
linear different sizes (10, 13, 17, 31, 37, and 71).
For the critical point: $J/kT_c = K_c = 0.113915$ is used as in ref.
[11].
The FORTRAN program used for simulations is listed below.

\bigskip

2. {\bf Problem}

Our problem here is to reexamine the universality scaling of 5D
Ising model with the susceptibility and the magnetization along lines
of constant $A = L^2(T-T_c)/T_c$.

\bigskip

3. {\bf Simulations and Results }

From our simulation for different sizes of lattice, by varying the
temperature near the critical temperature we get the ratio of susceptibility 
by dividing the susceptibility of temperature above $T_c$ to the susceptibility 
below $T_c$ at the same $|T-T_c|$. 
Then the ratio of susceptibility was drawn versus $|T_c-T|/T_c$ as shown
in figure (2) .

It can be seen that the ratio of susceptibility is roughly constant
for varying size of lattice but increases away from the critical
temperature.
When large lattice as $L=71$ is tested for different times (500 and 5000),
our simulation give the data as presented in figure 3.

This figure shows that the susceptibilities scatter much more than
the magnetizations. Now we test the universality of 5D Ising model
and vary $T$ along lines of constant $A = L^2(T-T_c)/T_c$ below, at and
above $T_c$ with different $L$ (10,13,17,23,31) for long time (500000).
Then the data are obtained as seen in figures 4 and 5.

Now if the average of the absolute value of magnetization is taken,
and plotted with the size of lattices with all constants $A=+1,0,-1$
with log-log plot , the slopes are obtained in figure 4, in
agreement with previous theories and simulations [1,8,9].

Then by drawing the susceptibility versus the size $L$ of lattices for
the constants $A=+1,0,-1$ with log-log plot , we get different
slopes, twice as large as for the magnetization in the previous
figure, as shown in figure 5.

\bigskip

4. {\bf Programming used in Simulations:}

A: Main program:

\small{
\begin{verbatim}
      PARAMETER(L=17,L2=L*L,L3=L2*L,L4=L2*L2,L5=L3*L2,
     1 LMAX=L5+2*L4)
      INTEGER *8 IBM,IEX
      DIMENSION IEX(-10:10)
      BYTE IS(LMAX)
      DATA TC,MAX,IBM,ISEED/0.113915,500000,1,1/
      IBM=2*ISEED -1
      CONST=0.0
      T=-(TC*CONST/L2)+TC
      T1=TC/T-CONST
C     T= TI*(1.0-0.1/(L*L))
      PRINT *,L,T,T1,MAX,ISEED
      LP1=L4+1
      L2PL=L5+L4
      DO 1 I=1,LMAX
1     IS(I)=1
      DO 2 IE=-10,10
      IBM=IBM*16807
      EX=EXP(-2.0*IE*T)
2     IEX(IE)=2147483648.0D0*(4.*EX/(1.0+EX)-2.0)*2147483648.0D0
      DO 3 MC=1,MAX
      DO 4 I=LP1,L2PL
      IE=IS(I)*(IS(I-1)+IS(I+1)+IS(I-L)+IS(I+L)+IS(I-L2)+IS(I+L2)
     1   +IS(I-L3)+IS(I+L3)+IS(I-L4)+IS(I+L4))
      IBM=IBM*16807
      IF (IBM.LT.IEX(IE)) IS(I)= -IS(I)
      IF(I.NE.2*(L4)+1) GOTO 4
      DO 7 J=1,L4
7     IS(J+L5+L4)=IS(J+L4)
4     CONTINUE
      FACTOR=1.0/(L*L*L*L*L)
      DO 5 I=1,L4
5     IS(I)=IS(I+L5)
      MAGN=0
      DO 6 I=LP1,L2PL
6     MAGN=MAGN+IS(I)
      X=MAGN*FACTOR
3     PRINT *,MC,MAGN,X
      STOP
      END
\end{verbatim} }

B: Analysis program:

\small{
\begin{verbatim}
      INTEGER*8 MAGN,SUMMAG,SUMSQU
      REAL*8 X, AVERGESUMMAG,AVERGESUMSQU
      READ *,L,T,T1,MAX,ISEED
      L5=L*L*L*L*L
      SUMMAG=0
      SUMSQU=0
      ISUMMAG=0
      DO 100 I=1,MAX
      READ *,MC,MAGN
      X=MAGN
      IF(MC.LE.(MAX/2)) GO TO 100
      SUMMAG=SUMMAG+X
      ISUMMAG=ISUMMAG+MAGN
      SUMSQU=SUMSQU+X*X
C     PRINT *, MC, ISUMMAG,ISUMSQU
100   CONTINUE
      AVERGESUMMAG=SUMMAG/(MAX*0.5D0)
      AVERGESUMSQU=SUMSQU/(MAX*0.5D0)
      X=AVERGESUMMAG/L5
      CHI=(AVERGESUMSQU-AVERGESUMMAG**2)/L5
      PRINT 1,L,T,X,CHI,ABS(T),X*X*CHI
1     FORMAT (1X,I2,5F15.5)
      STOP
      END
\end{verbatim} }

\bigskip

5. {\bf Conclusion}

We thus confirmed [1,8,9] that finite size scaling in high dimensions
is
described by different exponents if we approach the critical point
along different lines in the plane of $T-T_c$ versus $1/L^2$, above, at
and
below $T_c$. This holds is not only for the magnetization [8] but also
for the susceptibility, though the susceptibilities above $T_c$ are
problematic.

\bigskip

{\bf References}

\parindent 0pt

[1] X. S. Chen and V. Dohm, Int. J. Mod. Phys. C 9, 1073 (1998).
 
[2] W. Selke, L.N. Shchur, J. Phys. A 39 (2005) L739.
 
[3] M. Schulte, C. Drope, Int. J. Mod. Phys. C 16 (2005) 1217.
 
[4] M.A. Sumour, M.M. Shabat, Int. J. Mod. Phys. C 16 (2005) 585.
 
[5] M.A. Sumour, M.M. Shabat, D. Stauffer, the Islamic University
Journal (Series of Natural Studies and Engineering) Vol.14, No.1,
P.209-212, 2006.
 
[6] F.W.S. Lima, D. Stauffer, Physica A 359 (2005) 423.
 
[7] M.A. Sumour, M.M. Shabat, D. Stauffer, and A.H. El-Astal, Physica
A(2006) in press.
 
[8] M. Cheon, I Chang, and D. Stauffer, Int. J. Mod. Phys. C 10, 131
(1999).
 
[9] Z. Merdan et al, Physica A (2006),(in press).
 
[10] L. Zekri, Int. J. Mod. Phys. C 16,199 (2005).
 
[11] Luijten E. and Bl\"ote H. W. J., Phys. Rev. Lett., 76 (1996)
1557, 3662.
\end{document}